\documentclass[11pt]{article}

\usepackage{epsf}
\usepackage{rotating}
\usepackage{epsfig}
\usepackage{amssymb}  
\usepackage{amsmath}
\usepackage{psfrag}

\newcommand{\beq}{\begin{equation}} 
\newcommand{\eeq}{\end{equation}}
\newcommand{\be}{\begin{eqnarray}} 
\newcommand{\ee}{\end{eqnarray}}

\newcommand{\dd}{{\mathrm d}} 
\newcommand{\e}{{\mathrm e}}
\newcommand{\de}{\partial} 
\newcommand{\refeq}[1]{(\ref{#1})}

\newcommand{\vqm}{V_\mathrm{QM}}
\newcommand{\vdqm}{\tilde{V}_\mathrm{QM}}

\newcommand{\jost}{{\cal J}}

\psfrag{UQM}{$U$}
\psfrag{DUQM}{$\tilde{U}$}
\psfrag{Bruch}{\small $$}
\psfrag{c_ren}{\small $$}
\begin{document}
\begin{titlepage}

\vfill
\begin{flushright}
\today\\
QMW-PH-01-17\\
hep-th/0112150\\
\end{flushright}

\vskip 2cm
\begin{center}
{\bf \Large CFT/CFT interpolating RG flows and}
\vskip 0.3cm
{\bf \Large the holographic \boldmath {$c$}-function}

\vskip 1.6cm
{\large Dario Martelli and Andr\'e Miemiec}\\

\vskip 0.5cm
{\it
Department of Physics\\
Queen Mary College, University of London\\
Mile End Road, London E1 4NS, UK
}\\
\vspace{1cm}
\end{center}
\par

\begin{abstract}
\noindent
We consider holographic RG flows which interpolate 
between non-trivial ultra-violet (UV) and infra-red (IR)
conformal fixed points. We study  the ``superpotentials'' which characterize
different flows and discuss their expansions near the fixed points.
Then we focus on the holographic $c$-function as defined from
the two-point function of the stress-energy tensor. 
We point out that the equation for the metric fluctuations
in the background flow  
is equivalent to  a scattering problem and we use this
to obtain an expression for the $c$-function in terms of the 
associated Jost functions. 
We propose two explicit models that realize UV-IR flows.
In the first example 
we consider a thin wall separating two AdS spaces with different radii, 
while in the second one the UV region is replaced with an asymptotically
AdS space.
We find that the holographic $c$-function obeys the expected properties.
In particular it reduces to the correct -- nonzero -- 
central charge in the IR limit.   
\end{abstract}

\end{titlepage}

\section{Introduction}
\label{intro}

Since the work of Zamolodchikov \cite{Zamolodchikov:1986gt}, the proof 
of a $c$-theorem for quantum field theories in more than two dimension has
remained an open problem. In the last years the idea of holography, 
namely the extension of the AdS/CFT correspondence 
to non conformal theories, has provided new tools for addressing 
this issue. In this context, it was shown 
\cite{Girardello:1998pd,Freedman:1999gp} that 
in any dimension there exists a function of the gravity fields which,
under suitable physical assumptions, fulfills the basic requirements of a 
central function, {\em i.e.} a monotonic function of some energy scale, 
which in 
the IR and UV limits reduces to the central charges of the fixed point
conformal field theories \cite{Anselmi:1998rd}.
Since it is not obvious how observables on the field theory side are 
related to the supergravity fields, the field theoretical meaning of 
this function has remained unclear. 
In \cite{Anselmi:2000fu} a natural definition of a holographic $c$-function 
was proposed,
which is related to the OPE of the stress-energy tensor \cite{Anselmi:1998rd} 
as (here in $d=4$)
\be
\langle T_{\mu\nu}(x)T_{\rho\sigma}(0) \rangle & = &
-\frac{1}{48\pi^4}\Pi^{(2)}_{\mu\nu\rho\sigma}
\left[\frac{c(x)}{x^4}\right]+\pi_{\mu\nu}
\pi_{\rho\sigma}\left[\frac{f(x)}{x^4}\right] 
\ee 
where $\pi_{\mu\nu}=\de_\mu\de_\nu-\eta_{\mu\nu}\de^2$ and
$\Pi^{(2)}_{\mu\nu\rho\sigma}=2\pi_{\mu\nu}\pi_{\rho\sigma}-
3(\pi_{\mu\rho}\pi_{\nu\sigma}+\pi_{\mu\sigma}\pi_{\nu\rho})$.
Hence one can compute it in holographic flows by analyzing the fluctuations
of the transverse traceless (TT) graviton in a given
background \cite{Gubser:1998bc,witten}.
This proposal has been tested in a few cases where the flows are known 
explicitly, and for which the fluctuations can be expressed in closed form
\cite{Porrati:2000ha,Porrati:2001nb}.
However, these flows are all singular in the IR, describing non conformal 
theories and/or vacua (so-called Coulomb branch flows). Moreover, a proof 
that the suggested
$c$-function satisfies its characterizing requirements is still
missing.

In order to study the holographic central function further, one should 
construct explicitly background RG flows, which interpolate
between two non-trivial fixed point conformal field theories. By
``interpolating'' flows we mean domain wall solutions (not 
necessarily supersymmetric), which start at a critical point of the 
supergravity 
potential and end at a nearby critical point. Even though there are a few
cases in which such flows are known implicitly
\cite{Girardello:1998pd,Distler:1998gb,Freedman:1999gp},
it is somewhat disappointing that no analytic solutions have 
been found so far.

A  useful method for finding
supergravity solutions is to rewrite the equations of motion as a first order
system, supplemented with a non-linear equation for a
``superpotential'' (see {\em e.g.} \cite{DeWolfe:2000cp,Campos:2000yu}). 
Experience has shown that when such a ``superpotential'' can be written down 
explicitly this is an unmistakable sign of underlying supersymmetry. 
Nevertheless, one is not committed to restrict to such a class of solutions, and in
fact the possible solutions that arise from a given supergravity potential
are far more richer \cite{Campos:2000yu}. It turns out that there are infinite 
solutions which represent deformations of a UV conformal field theory.
Two of them are somewhat special and correspond to flows dual to
operators acquiring a vacuum expectation value (Coulomb branch flows) and the
interpolating flow.  In this paper we study global and local properties
of the ``superpotential'', in order to characterize the interpolating
solutions.     

The problem of solving the TT graviton fluctuation in a Poincar\'e invariant
flow leads generically to a Schr\"odinger equation with a 
potential of supersymmetric type \cite{Freedman:2000gk}
which depends non trivially on the original supergravity potential.
We point out that for an interpolating RG flow the problem is 
that of a partial wave scattering off a central potential.
This allows to study properties of the fluctuations without
detailed knowledge of the flow and to obtain some informations about 
the associated holographic $c$-function.
A general proof that the $c$-function satisfies the expected properties 
still remains an open problem. 
Nevertheless, we have checked that these properties are
respected in some simplified models, where the background flow is obtained
by patching together space-times on which the fluctuation can be solved in
closed form.
Some properties of two-point functions in generic interpolating flows were 
studied in \cite{Porrati:1999ew}. 
Previous studies of Schr\"odinger potentials associated with 
the TT graviton fluctuations appeared for instance in 
\cite{Freedman:2000gk,Cvetic:2000xx,Bakas:1999fa}. 
However, in these cases, the analysis was devoted to IR-singular 
supersymmetric flows. 

The paper is organized as follows. In section \ref{back}
we give a detailed analysis
of solutions related to a given potential in a one-scalar
model of the holographic RG. We study the global structure of the 
``superpotential'' and write out the  
local expansion of some particular solutions near the fixed points.
In section \ref{scatt} we move to the analysis of the fluctuations and the associated quantum mechanical potential.
We write down an expression for the holographic $c$-function in terms of Jost
functions of a suitable scattering problem, and obtain
some general properties consistent with field theory expectations.
Finally, in section \ref{astutissimi}, 
we study the $c$-functions related to two examples
which model the flows with piecewise simple space-times. This allows us to test the general prescription
for computing correlation functions in IR-conformal flows.

\section{Background flows}
\label{back}

We begin with a review of holographic RG flows, explaining in particular 
how different trajectories arise from a given supergravity potential.
The starting point is an action in $d+1$ dimensions\footnote{Our convention for
the metric signature is $\eta_{ij} = (-,+,\cdots,+)$. And we have set the Newton
constant $\kappa=1$.}
\be
S = \int d^{d+1}x\sqrt{g} \Big(R -\frac{1}{2} (\de\phi)^2-V(\phi)\Big)-2\int d^d x
\sqrt{\hat{g}} K
\label{start}
\ee
which one can think of as a truncation of some supergravity action.
We consider the Poincar\'e invariant ansatz for the background
flow and work from the beginning in
the conformal gauge 
\be
\dd s^2 & = & \e^{2A(z)}(\dd z^2+ \eta_{ij}\dd x^i \dd x^j)\\
\phi & = & \phi (z)~.
\ee
We want to address trajectories that interpolate between two AdS fixed
points, therefore we shall assume that the potential possesses two nearby
extrema, and so $V$ must be negative definite.
The second order equations of motion resulting from \refeq{start} are
equivalent to the following first order equations 
\be
\label{BPS}
\frac{d A}{d z} ~=~  \frac{\e^{A}}{2(d-1)}\,W\qquad\qquad\qquad
\frac{d \phi}{d z} ~=~  -\,\e^{A}\,W_\phi~,\\[-0.5ex] 
\nonumber
\ee
with the ``superpotential'' $W$ defined by the equation
\be
V  & = &\frac{1}{2}\,W_\phi^2 ~-~ \frac{d}{4(d-1)}\,W^2~.
\label{nonlin}
\ee
For any given solution of \refeq{nonlin} there is a mirror one, which has
opposite overall sign. We will restrict the analysis to those which are 
negative definite.

\subsection{Solutions for the superpotentials}

The non-linear nature of \refeq{nonlin} makes the analysis of possible 
trajectories in the phase space ($W$-$\phi$ plane) non-trivial. In spite of
being a first order equation, it is not true that in general a one-parameter
family of solutions covers the full phase space. 
In this respect, the existence of
{\em supersymmetric} solutions is a remarkable example of isolated
trajectories. These solutions arise, generically, by
requiring vanishing dilatino and gravitino variations, and are therefore the
true superpotentials of the theory.

In order to discuss the different kind of solutions we have solved numerically
\refeq{nonlin} for some potential which posses one maximum (UV fixed point)
and one minimum (IR fixed point).
In Fig. \ref{plotcubic} it is depicted a typical potential.\\ 
%%%%%%%%%%%%%%%%%%%%%%%%%%%%%%%%%%%%%%%%%%%%%%%%%
\parbox{\textwidth}{\vspace{0.2cm}
%%%%%%%%%%%%%%%%%%%%%%%%%%%%%%%%%%%%%%%%%%%%%%%%%
  \refstepcounter{figure}
  \label{plotcubic}
  \begin{center}
  \begin{turn}{0}
  \makebox[5cm]{
     \epsfxsize=5cm
     \epsfysize=4cm
     \epsfbox{cubicpotential.epsi}
     }
  \end{turn}
  \end{center}
  \center{{\bf Fig.{\thefigure}.} The cubic potential
  $V=-12+1/2m^2\phi^2+1/3m^2\phi^3$ with $m^2=-3.8$.}
%%%%%%%%%%%%%%%%%%%%%%%%%%%%%%%%%%%%%%%%%%%%%%%%%
}\\
%%%%%%%%%%%%%%%%%%%%%%%%%%%%%%%%%%%%%%%%%%%%%%%%%

\noindent
As shown in Fig.~\ref{phasespace}, the phase space is bounded from above
by the curve \linebreak $V(\phi)+\frac{d}{4(d-1)}W^2_\mathrm{up}=0$.
The two critical points of $V$ correspond
to critical points in the phase space and it turns out that the IR one is
repulsive, while the UV one is attractive. There is
a continuum of solutions which approach the UV fixed point, whereas only one
originates from the IR fixed point.\\ 
%%%%%%%%%%%%%%%%%%%%%%%%%%%%%%%%%%%%%%%%%%%%%%%%%
\parbox{\textwidth}{\vspace{0.2cm}
%%%%%%%%%%%%%%%%%%%%%%%%%%%%%%%%%%%%%%%%%%%%%%%%%
\parbox{0.33\textwidth}
{
  \refstepcounter{figure}
  \label{phasespace}
  \begin{center}
  \begin{turn}{0}
  \makebox[4cm]{
     \epsfxsize=4.5cm
     \epsfysize=3.5cm
     \epsfbox{w.epsi}
     }
  \end{turn}
  \end{center}
  \center{{\bf Fig.{\thefigure}.} $W$}
}\hfill
\parbox{0.33\textwidth}
{
  \refstepcounter{figure}
  \label{phasespace1}
  \begin{center}
  \begin{turn}{0}
  \makebox[4cm]{
     \epsfxsize=4.5cm
     \epsfysize=3.5cm
     \epsfbox{wprime.epsi}
  }
  \end{turn}
  \end{center}
  \center{{\bf Fig.{\thefigure}.} $W_\phi$}
}
\parbox{0.33\textwidth}
{
  \refstepcounter{figure}
  \label{phasespace2}
  \begin{center}
  \begin{turn}{0}
  \makebox[4cm]{
     \epsfxsize=4.5cm
     \epsfysize=3.5cm
     \epsfbox{wsecond.epsi}
  }
  \end{turn}
  \end{center}
  \center{{\bf Fig.{\thefigure}.} $W_{\phi\phi}$}
}
\center{Solutions corresponding to the potential of Fig. \ref{plotcubic}. The
dashed curves correspond to  the interpolating and ``Coulomb branch''
solutions.}
%%%%%%%%%%%%%%%%%%%%%%%%%%%%%%%%%%%%%%%%%%%%%%%%%
}\\
%%%%%%%%%%%%%%%%%%%%%%%%%%%%%%%%%%%%%%%%%%%%%%%%%

\noindent
There are two distinguished curves (dashed lines in the Figure):
One is the interpolating trajectory, the other a lower bound delimiting
the solutions 
flowing to the UV point. The distinguished role played by these two solutions
is better understood by looking at the first and second derivatives, shown in
Fig.~\ref{phasespace1} and \ref{phasespace2}. 
In particular, notice that
for all the solutions ending at the UV fixed point (including the interpolating one) the second derivative
$W_0^{(2)}=\Delta-d\approx -1.55$ at the UV point, whereas there is only one
solution such that $W_0^{(2)}=-\Delta\approx -2.45$. $\Delta$ is the conformal
dimension of the dual CFT operator, which is related to the mass of the
scalar field by the usual formula $\Delta = (d+\sqrt{d^2+4m^2R^2})/2\, $.
The latter solution coincides usually with the supersymmetric superpotential
when supersymmetry is preserved \cite{Campos:2000yu,Skenderis:1999mm}.  
Notice that the solutions not flowing to the UV point must end at the 
upper bound of the allowed phase space and are all singular there, as can be seen
from the second derivatives blowing up. \\ 
%%%%%%%%%%%%%%%%%%%%%%%%%%%%%%%%%%%%%%%%%%%%%%%%%
\parbox{\textwidth}{\vspace{0.2cm}
%%%%%%%%%%%%%%%%%%%%%%%%%%%%%%%%%%%%%%%%%%%%%%%%%
\parbox{0.33\textwidth}
{
  \refstepcounter{figure}
  \label{nonbfphasespace}
  \begin{center}
  \begin{turn}{0}
  \makebox[4cm]{
     \epsfxsize=4.5cm
     \epsfysize=3.5cm
     \epsfbox{nonbfw.epsi}
     }
  \end{turn}
  \end{center}
  \center{{\bf Fig.{\thefigure}.} $W$}
}\hfill
\parbox{0.33\textwidth}
{
  \refstepcounter{figure}
  %\label{}
  \begin{center}
  \begin{turn}{0}
  \makebox[4cm]{
     \epsfxsize=4.5cm
     \epsfysize=3.5cm
     \epsfbox{nonbfwprime.epsi}
  }
  \end{turn}
  \end{center}
  \center{{\bf Fig.{\thefigure}.} $W_\phi$}
}
\parbox{0.33\textwidth}
{
  \refstepcounter{figure}
  %\label{}
  \begin{center}
  \begin{turn}{0}
  \makebox[4cm]{
     \epsfxsize=4.5cm
     \epsfysize=3.5cm
     \epsfbox{nonbfwsecond.epsi}
  }
  \end{turn}
  \end{center}
  \center{{\bf Fig.{\thefigure}.} $W_{\phi\phi}$}
}
\center{Solutions for $W$ for a mass $m^2=-20$ violating the BF bound.
The dashed curves correspond to the solution originating in the IR fixed point.}
%%%%%%%%%%%%%%%%%%%%%%%%%%%%%%%%%%%%%%%%%%%%%%%%%
}\\
%%%%%%%%%%%%%%%%%%%%%%%%%%%%%%%%%%%%%%%%%%%%%%%%%

\noindent
Before going to the discussion of the local expansions near the fixed points,
let us remark on the issue of stability. In view of the fact that there
exist more
solutions flowing to the UV point than only the two found by Taylor
expansion (see below), one could ask whether some of these solutions do exist 
even for scalar modes violating the Breitenlohner--Freedman (BF) 
bound $m^2R^2<-d^2/4$ \cite{Breitenlohner:1982bm}.
In fact, the numerical solutions
show that in this case there are no trajectories flowing to the UV point, as
depicted in Fig. \ref{nonbfphasespace}. 
Therefore the condition for stability discussed in \cite{Townsend:1984iu} 
is valid beyond the local Taylor expansion.

The situation for a mass
precisely at the BF bound is the same as in the BF-stable case. There 
is still a continuum of solutions ending in the UV fixed point, all of them 
having the same second derivative $W_0^{(2)}=\Delta-d=-\Delta$.

\subsection{Expansion of the superpotentials near the fixed points}
\label{superpexp}

We have seen that the interpolating solution should exists in general.
We will proceed now with the local analysis of $W$ in the vicinity of the AdS
fixed points, which are by definition critical points of the gravity potential
$V$, so elucidating the meaning of the two distinguished solutions.
In general we do not need to specify whether we are near an IR or UV fixed
point. Near a critical point $\phi_0$, we write the gravity potential as
\be
V (\Phi) & = & -\frac{d(d-1)}{R^2} + \frac{1}{2}m^2\Phi^2+\frac{1}{3!}
V^{(3)}\Phi^3+\cdots~
\ee 
where $\Phi=\phi-\phi_0$, and we have expressed the constant term in terms
of the AdS radius $R$.
Let us assume that a Taylor expansion for $W$ near $\phi_0$ holds,
and expand equation \refeq{nonlin} in power series, using the ansatz 
\be
W (\Phi) & = & W_0 + W^{(1)}_0\Phi+\frac{1}{2}W^{(2)}_0\Phi^2+
\frac{1}{3!}W^{(3)}_0\Phi^3+\cdots~.
\ee

If $W_0^{(1)}\neq 0$, the resulting zero-th order equation
contains the two undetermined coefficients  $W_0$ and $W_0^{(1)}$, one of which, say $W_0$,
can be taken as an integration constant. Equating the linear terms gives
$W_0^{(2)} = d/(2d-2)~W_0\,$,
showing that $W_0^{(2)}$ is always negative, which for instance is 
observed in Fig. \ref{phasespace2}. Higher order equations are
algebraic\footnote{There is no obstruction in solving these equations, contrary
to the $W^{(1)}_0=0$ case.}
equations for $W_0^{(n)}$ in terms of the Taylor coefficients of $V$,
which parametrically depend on $W_0$. 

If $W^{(1)}_0\!=0$, the expansion of \refeq{nonlin}
is substantially changed, and needs to be analyzed separately. 
Since in this case a critical point of $W$ does coincide with one
of $V$, we can write the more familiar expansion \cite{Kalkkinen:2001vg} 
in terms of the conformal dimension $\Delta$ 
\be
W (\Phi) & = & -\frac{2(d-1)}{R} + \frac{\Delta-d}{2R}\Phi^2
+\frac{1}{3!}\frac{RV_0^{(3)}}{3\Delta-2d}\Phi^3+\cdots~.
\label{Wdef}
\ee
Notice that $\Delta$ is usually thought of as pertinent to the UV fixed
point, in which case $0>m^2R>-d^2/4$ and the conformal dimension 
ranges accordingly in
\be
d > \Delta_\mathrm{UV} \geq d/2~.
\label{boundUV}
\ee
In the IR instead, the potential is at a local minimum, thus $m^2>0$, and
in this case the conformal dimension has no upper bound, satisfying in general
only
\be
\Delta_\mathrm{IR} > d~.
\label{boundIR} 
\ee  
The second order in the Taylor expansion is quadratic in $W_0^{(2)}$, therefore
there exists another solution which gives rise to a superpotential
\be
{\hat W} (\Phi) & = & -\frac{2(d-1)}{R} 
-\frac{\Delta}{2R}\Phi^2-\frac{1}{3!}\frac{RV_0^{(3)}}{3\Delta-d}
\Phi^3+\cdots~.
\label{Wvev}
\ee
The expansion \refeq{Wvev} corresponds clearly to a Coulomb branch flow. 
The interpolating solution instead, corresponds to a true deformation. Some 
numerical analysis has given evidence that this is actually the solution
which admits a Taylor expansion, though its radius of convergence is not 
known  in general. The higher level equations of the expansion take the 
following form
\be
[n\Delta -(n-1)d]\,W_0^{(n)} & = & R V_0^{(n)} + 
P(\Delta,V_0^{(3)},\dots,V_0^{(n-1)})~,
\label{special:delta}
\ee
where $P$ is a polynomial in the Taylor coefficients of $V$.
Thus, we find that there is a series of operators with rational
dimensions for which the Taylor expansion of $W$ explicitly 
breaks down. The expansion
does not break down if the left hand side of \refeq{special:delta}
vanishes identically. This happens for instance for the flow found in 
\cite{Girardello:2000bd}, where a supersymmetric (and analytic) superpotential
exists, despite the operator has dimension $\Delta = 3$. 
In general the impossibility of performing a Taylor expansion
does not imply that no solution exists. In fact, we find that insertion 
of a logarithmic term is enough to solve the equation, at least locally.
This solution reads
\be
W (\Phi) = -\frac{2(d-1)}{R} + \frac{\Delta-d}{2R}\Phi^2+\cdots
+\frac{1}{n!}W_0^{(n)}\Phi^n+\frac{1}{n!}X_0^{(n)}\Phi^n\log\Phi+\cdots~.
\label{Wlog}
\ee
Higher order terms contain powers of logs and all the coefficients of the 
series are fixed by the differential equation \refeq{nonlin}, except
$W_0^{(n)}$, which remains as a new integration constant. 
The structure of this expression is formally
similar to the expansion that arises in solving the Einstein equations
near an AdS fixed point \cite{fefferman}. In that case a logarithmic 
term appears in the metric at order $d$, and for the scalar field at order
$n=2\Delta-d$, whereas here we find that it occurs for operators of dimension
\be
\Delta & = & \frac{n-1}{n}\,d
\label{fractop}
\ee
where $n$ is an integer $n\geq 3$. Both types of operators contribute 
to the conformal anomaly in the fixed point CFT 
\cite{Petkou:1999fv,Kalkkinen:2001vg}. We have shifted a discussion
of the above expansion and the special conformal dimensions \refeq{fractop} 
to Appendix \ref{reno}, since it is not needed for what follows.

\subsection{Expansion of the fields near the fixed points}

Now we turn to the first order equations for the fields, and solve them
perturbatively, near a fixed point. 
From \refeq{BPS} one gets 
\be
\frac{dA}{d\phi} & = & -\,\frac{1}{2(d-1)}\, \frac{W}{W_\phi}
\ee
which can be solved for $A$ upon expanding the right hand side near the fixed
point. Afterwards, integrating order by order, one finds a series for $z$ in
terms of $\phi$. This series can be inverted, the leading term being linear,
so that eventually one can read off the asymptotic expansions for $A(z)$ and
$\phi(z)$. It turns out that the following expansions hold
\be
\phi (z) & = & \phi_0+\tau \Big[1+c_1\tau+c_2\tau^2+
{\cal O}(\tau^3)\Big]\label{asymphi}\\
\e^{A(z)} & = & \frac{R}{z-z_0}\Big[1+a_2\tau^2+a_3\tau^3+
{\cal O}(\tau^4)\Big]~,\label{asymA}
\ee
where
\be
\tau & \equiv & \left(\frac{z-z_0}{R\e^{-A_0}}\right)^{d-\Delta}~.
\label{deftau}
\ee
The first few coefficients of the expansions can be found in Appendix
\ref{coefficients}\footnote{Notice that the linear coefficient in the
expansion of $\e^{A}$ vanishes.}. 
The expansions are sensible both in the UV and IR
limits, as the sign of $d-\Delta$ flips appropriately in the two cases.
The two integration constants appear in a substantially trivial way:
$z_0$ is a shift of $z$ and $A_0$ a rescaling. We will omit them in the
following, with the proviso that one should not
prescribe them simultaneously in UV and IR regions. 
In the UV a similar expansion holds using the hatted
superpotential \refeq{Wvev}, with the change that the exponent in 
\refeq{deftau} is now $\Delta$.  

From the above expansions we can confirm that 
$\hat W$ produces a Coulomb branch flow, as the leading behavior of the 
scalar field for small $z$ is $\Phi (z) \sim z^\Delta$~.
Conversely, for a superpotential of type \refeq{Wdef}, 
the leading behavior of \refeq{asymphi} is
$\Phi (z) \sim z^{d-\Delta}$~,
which is a deformation flow. Moreover, a closer look to 
\refeq{asymphi} also shows that at subleading order the signature of 
a vev never appears. In fact this would be the case if for some integer $n$, 
$(d-\Delta) n =  \Delta $, 
but we have excluded these cases from the analysis, in that they led to a
superpotential modified with the logarithmic term. We therefore see 
that the superpotential generates a pure deformation flow, with vanishing
vacuum expectation value for the operator $\langle {\cal O}_\Delta\rangle=0$. 

We would like to stress that the expansions \refeq{asymphi} and \refeq{asymA}
are slightly different from the usual Fefferman--Graham type expansions 
encountered in the literature \cite{fefferman,deHaro:2001xn,Bianchi:2001de}.
In fact these are series in $z^{d-\Delta}$, whose exponent is in general 
{\em not} an integer. Apparently this fact has not been noticed before, since
all flows extensively studied in the literature have integer exponents.
The fact that there are no       
logarithmic terms is not in contradiction with the usual   
expansions since we are dealing with a Poincar\'e invariant background flow,
for which those terms actually vanish.

\section{The holographic $c$-function from scattering} 
\label{scatt}

In this section we will consider the holographic $c$-function proposed  
in \cite{Anselmi:2000fu} following field theoretical motivations.
It can be computed as the Fourier transform of the transverse traceless (TT)
stress-energy correlation function, related to the ``flux''-factor initially
discussed \cite{Gubser:1998bc,witten} in the pure AdS/CFT case.
In particular the $c$-function is related to the flux
as
\be
c(x) & = & (2\pi)^\frac{d}{2} x^{\frac{3}{2}d-3}\int_0^\infty
d q~ q^{\frac{d}{2}-4}{\cal F}(q)J_{\frac{d}{2}-1}(qx)~
\label{hankeltransf}
\ee 
where $x=\sqrt{x_i x^i}$ measures distances in the boundary theory.
The flux is obtained from the linearized fluctuations of the 
TT graviton in the background of an interpolating flow. 
By a slight redefinition of the metric fluctuations $h_{ij}^{TT}$, 
\begin{eqnarray}
   h_{ij}^{TT} ~=~ \e^{ikx}\,\chi(z)\,\xi_{ij} (x) 
   \qquad & \mathrm{and} &\qquad
   \chi (z)    ~=~ \e^{-\frac{d-1}{2}A(z)}\,\psi (z)~,
\end{eqnarray}
the linearized equation of motion can be casted \cite{Freedman:2000gk} 
into a Schr\"odinger equation 
\begin{eqnarray}\label{schr1}
   -\,\frac{d^2}{dz^2}\,\psi (z)~+~\left[\,\vqm (z)\,-\,k^2\,\right]\,\psi (z)
    & = & 0~.
\end{eqnarray}
The quantum mechanical potential is of supersymmetric type and can be 
written as
\begin{eqnarray}\label{vqm}
   \vqm  ~=~  W_\mathrm{QM}^2\,+\,\frac{dW_\mathrm{QM}}{dz}
              \qquad &\mathrm{with}& \qquad 
              W_\mathrm{QM} ~\equiv~ \frac{d-1}{2}
              \frac{dA}{dz}~.
\end{eqnarray}
There exists also a ``supersymmetric partner potential'' $\vdqm$, whose 
properties are closely related to $\vqm$, which is obtained simply by 
$W_\mathrm{QM}\to -W_\mathrm{QM}$. One can 
express the potentials in terms of the supergravity quantities by 
\begin{eqnarray}\label{sugratoqm}
   \vqm ~=~ -\frac{\e^{2A}}{16}\left( W^2+8V\right)
   \qquad 
   \vdqm  ~=~ \frac{\e^{2A}}{16}\left( 3W^2+8V\right)~. 
%   \label{sugratodualqm}
\end{eqnarray}
These expressions do not depend on $d$ and one can easily 
verify that $\vdqm$ is positive definite as long as $d>2$.

We want to determine the type of quantum mechanical potential arising 
from an interpolating flow and focus on its local properties 
near the origin and infinity. The information about the 
potential in these regimes, together with the positivity  of 
$\vdqm$ mentioned above, gives a handle to apply scattering theory 
to the problem in order to obtain some insight into the central function 
associated with the flow.

Substituting the expansions \refeq{asymphi} and \refeq{asymA} into 
\refeq{sugratoqm}, the resulting expansion for the potential reads
\begin{eqnarray}\label{expan:vqm}
   \vqm (z) &=& \frac{\lambda^2-1/4}{z^2}
            ~+~ \frac{1}{z^2}\,
                \Big[\,
                        v_2\,\tau^2~+~ {\cal O}(\tau^3)\,
                \Big]
\end{eqnarray}
where we have introduced $\lambda=d/2$. Notice that the leading term is
independent of 
$R$ and has the form of a centrifugal barrier in the radial 
Schr\"odinger equation for scattering off a central potential with 
angular momentum $l\,=\,(d-1)/2$. The ``superpartner'' $\vdqm$ 
has a similar expansion but with $l$ replaced by $\tilde l= (d-3)/2$. 
The first non-zero coefficient of the expansion is given in the Appendix
\ref{coefficients}.
Since the centrifugal term in this expansion is the same in the UV and 
the IR limits, we split the potential  in  
\begin{eqnarray}
    \vqm (z) \equiv \frac{\lambda^2-1/4}{z^2} ~+~ U(z)
\end{eqnarray}
and consider $U(z)$ as the central potential that generates the scattering
of a particle in ordinary three dimensional space. From \refeq{expan:vqm} it
also follows that the leading behavior of $U(z)$ is given by 
\begin{eqnarray}
    z\to 0~~:\quad
    \quad
    U(z) ~\sim~ \frac{1}{z^{2-\kappa_\mathrm{UV}}} 
    \quad\quad 
    0 < \kappa_\mathrm{UV} ~=~ 2(d-\Delta_\mathrm{UV}) < d
    \label{UVbehav}\\
    z \to \infty~:\quad  
    \quad 
    U(z) ~\sim~ \frac{1}{z^{2+\kappa_\mathrm{IR\;}}} 
    \quad\quad
    0 \,<\, \kappa_\mathrm{IR} ~=~ 2(\Delta_\mathrm{IR}-d)\hspace{5ex}
    \label{longrange}
\end{eqnarray}
%%%%%%%%%%%%%%%%%%%%%%%%%%%%%%%%%%%%%%%%%%%%%%%%%
\parbox{\textwidth}{\vspace{0.2cm}
%%%%%%%%%%%%%%%%%%%%%%%%%%%%%%%%%%%%%%%%%%%%%%%%%
\parbox{0.5\textwidth}
{
  \refstepcounter{figure}
  \label{figureUQMd4}
  \begin{center}
  \begin{turn}{0}
  \makebox[4cm]{
     \epsfxsize=5cm
     \epsfysize=4cm
     \epsfbox{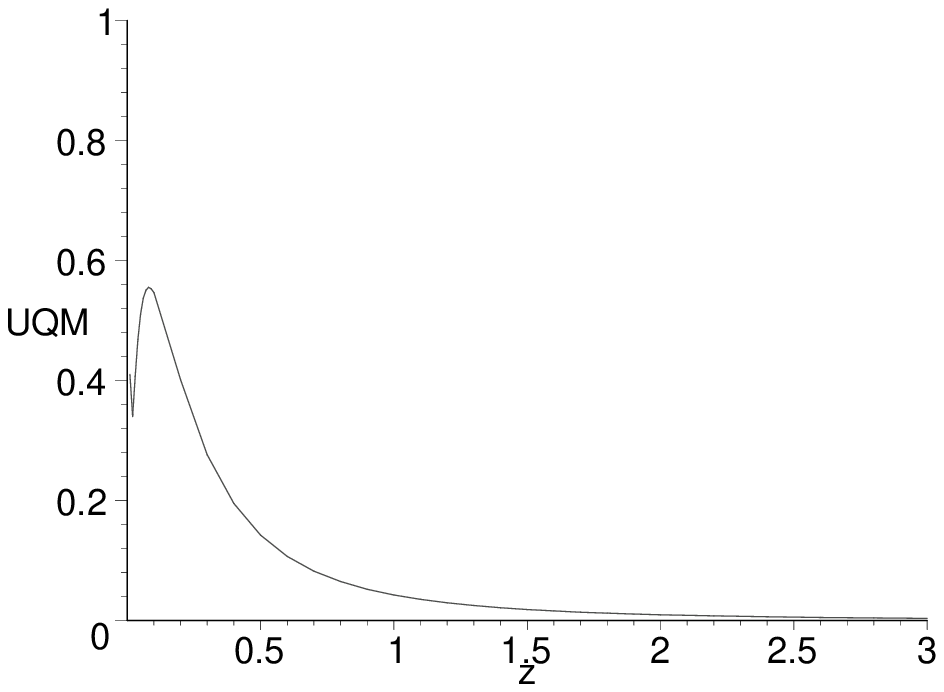}
  }
  \end{turn}
  \end{center}
  \center{{\bf Fig.{\thefigure}.}}
}\hfill
\parbox{0.5\textwidth}
{
  \refstepcounter{figure}
  \label{figureDUQMd4}
  \begin{center}
  \begin{turn}{0}
  \makebox[4cm]{
     \epsfxsize=5cm
     \epsfysize=4cm
     \epsfbox{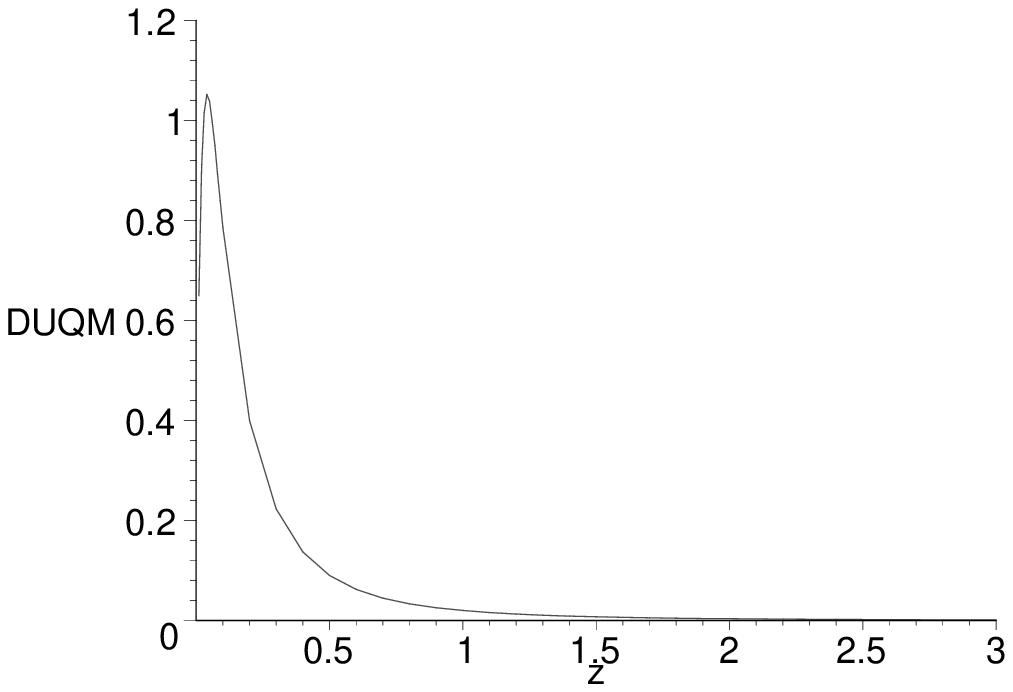}
  }
  \end{turn}
  \end{center}
  \center{{\bf Fig.{\thefigure}.}}
}
%%%%%%%%%%%%%%%%%%%%%%%%%%%%%%%%%%%%%%%%%%%%%%%%%
\center{Central potentials associated with a flow with 
$\kappa_\mathrm{UV}>2$.}
}\\
%%%%%%%%%%%%%%%%%%%%%%%%%%%%%%%%%%%%%%%%%%%%%%%%%
Moreover $\e^{2A(z)}$ is a non-singular, monotonically decreasing function,
and $W^2+8V$ is bounded along the flow, so that $U(z)$ is an 
acceptable potential in scattering theory (see for instance \cite{regge}). 
In Fig.~\ref{figureUQMd4} and Fig.~\ref{figureDUQMd4} we show the numerical
results for the quantum 
mechanical potentials $U$ and $\tilde U$ generated by the cubic potential 
of Fig.~\ref{plotcubic}.

We now turn to the computation of the correlation function.
The appropriate linear combination of independent solutions of the 
Schr\"odinger equation is selected by requiring regularity at the IR horizon. 
The solution exponentially suppressed in the Euclidean ($q=i|k|$) 
is the so-called Jost solution $\mathrm{f}(\lambda,-q,z)$, and using 
\refeq{newbasis} this can be related to the non-singular and the 
singular solutions $\varphi(\pm\lambda,q,z)$, whose behavior near 
the origin is given by \refeq{defreg}. If $d$ is even (which is the case of 
interest) the singular solution contains a subleading logarithmic
dependence on $z$, which must be taken into account 
in computing the correlation function\footnote
%%%%%%%%%%%%%%%%%%%%%%%%%%%%%%%%%%%%%%%%%%%%%%%%%%%%%%%%%%%%%%%%%%%%%%%
{
 We must use the solutions of the eq.~(\ref{schr1}). As long as 
 $\Delta$ is an {\em integer} the general theory of differential 
 equations tells us that the solutions defined at zero have the following 
 structure
 \begin{equation}
     u_1(z) ~=~ z^{\lambda+1/2}\,f_1(z)~, \qquad\quad
     u_2(z) ~=~ z^{-\lambda+1/2}\,f_2(z) \,+\,\log(z)\,u_1(z)\nonumber
 \label{andreansatz}   
 \end{equation}
 with $f_1(z)$ and $f_2(z)$ analytic functions and the logarithmic 
 contribution in the second solution  appears because $\lambda$ is 
 an integer. If $\Delta$ is irrational, the above structure of the solutions
 is not valid any more, and one would have to analyse the problem on a case by
 case basis. However, if $d-\Delta=m/n$ with integers $m,n$ ({\it i.e.}
 $\Delta$ is a rational number) by the change of variables $y=z^{1/n}$ one
 obtains an expansion of the above type, with $f_1(z)$ and $f_2(z)$
 series in $z^{1/n}$, and the main arguments of the section go trough.   
}.
%%%%%%%%%%%%%%%%%%%%%%%%%%%%%%%%%%%%%%%%%%%%%%%%%%%%%%%%%%%%%%%%%%%%%%%
In the notation of Appendix \ref{complexj} and by using the 
$\epsilon$-prescription for fixing the normalization \cite{Freedman:1999tz} 
the solution we are interested in reads 
\begin{eqnarray}
   \chi^\epsilon (q,z) 
   = \left(\,
                \frac{\e^{-A(z)}}{\e^{-A(\epsilon)}}\,
       \right)^{\lambda-1/2}\,
       \frac{
              \jost( \lambda,-q)\,\varphi(-\lambda,-q,z) - 
              \jost(-\lambda,-q)\,\varphi( \lambda,-q,z)
            }
            { 
              \jost( \lambda,-q)\,\varphi(-\lambda,-q,\epsilon) -
              \jost(-\lambda,-q)\,\varphi( \lambda,-q,\epsilon)
            }
\end{eqnarray} 
with $\jost(\pm\lambda,-q)$ the Jost functions characteristic to the 
potential $U(z)$.
This expression can now be used for computing the flux, as originally
explained in \cite{Gubser:1998bc,witten}, and further developed in
\cite{DeWolfe:2000xi,Arutyunov:2000rq,Bianchi:2001de,Muck:2001cy} for more 
general flows. We obtain the following expression
\begin{eqnarray}\label{firstresult}
{\cal F}(q) & = &  {\cal F}_\mathrm{log}(q) ~+~                            
                  \underbrace{
                               -\,2\,\lambda\, R_\mathrm{UV}^{2\lambda-1}\,
                                \frac{\jost( -\lambda,-q)}{\jost(\lambda,-q)}
                             }_{
                                 {\cal F}_\mathrm{int}(q)
                             }~.
\end{eqnarray}
It should be mentioned that the first term is closely related to the behavior
of the singular wave function at zero and contains in particular the 
contribution of the sub-leading logarithmic dependence. 
If $U$ vanishes at the origin ($\kappa_\mathrm{UV}>2$), the singular solution 
behaves as the $K_{\lambda}(q\,z)$ Bessel function and ${\cal F}_\mathrm{log}$
equals the pure 
AdS result. The second term ($\,{\mathcal{F}}_{\rm int}\,$) describes the
interaction with the potential and is generic.

Notice that we have included the normalization given by the UV AdS
radius, with the correct power (see {\em e.g.} \cite{Muck:2001cy} for a 
careful inclusion of the AdS radius in the correlation functions).
While in IR-singular flows one can usually 
rescale the overall normalization as it is more convenient, this is
strictly forbidden when considering interpolating flows, as in the IR
limit an independent scale appears.

The Jost functions of the two quantum mechanical potentials $\vqm$ and 
$\vdqm$ are in general related \cite{superjost} as
\be
\frac{\jost (\lambda,k)}{\tilde\jost (\lambda-1,k)} & = &
\frac{k}{k-i\gamma_0}~,
\ee 
where $\gamma_0^2$ is the ground state energy of the potential $\vdqm$.
This means that they have the same Jost functions, up to the addition of a 
bound state. In the present case $\vdqm$ is positive 
definite, hence it has no bound states and the continuum spectrum starts at 
zero energy. Therefore the Jost functions of the two potentials must be 
equal and in particular  $\jost (\lambda,-q)$ has no zeros \cite{regge}. 
It thus follows that in general ${\cal F}(q)$ has no massive poles
in $q$, which means that the dual field theory does not have a mass gap
or a spectrum of glue-balls, as expected from
conformal invariance being restored in the IR.   

In order to estimate the leading power terms of the $c$-function, we 
can extract its high and low energy behaviour by
performing a Born type expansion of the flux factor ${\cal F}_\mathrm{int}$
in \refeq{firstresult}. As recalled in Appendix \ref{complexj}, the Jost 
functions
admit a Born series expansion 
(~cf. eq.~\eqref{bornjost} \& eq.~\eqref{jost1}~) 
which follow from related integral equations\footnote
%%%%%%%%%%%%%%%%%%%%%%%%%%%%%%%%%%%%%%%%%%%%%%%%%%%%%%%%%%%%%%%%%%%%%%%%%
{We should mention
that the relevant integral equation is singular near the origin. For instance,
the integral in \refeq{estimate} is divergent at zero. However, assuming
that ultimately one can extract the finite contribution, the behaviour in
$q$ is determined by a simple scaling argument.
}. 
%%%%%%%%%%%%%%%%%%%%%%%%%%%%%%%%%%%%%%%%%%%%%%%%%%%%%%%%%%%%%%%%%%%%%%%%%
Accordingly, one can expand
the flux as ${\cal F}_\mathrm{int}(q)\,=\,\sum_{n=1}^\infty {\cal F}_n (q)$~,
where the first term reads
\begin{eqnarray}
 {\cal F}_1 (q) &=& -\,2\,\lambda\,R_\mathrm{UV}^{2\lambda-1}\,
                     \frac{\jost_1 (-\lambda,-q)}{\jost_0 (\lambda,-q)}
                     \nonumber\\ 
                &=& -\,\frac{4\,\lambda}{\Gamma(\lambda)\,\Gamma(\lambda+1)}\, 
                      R_\mathrm{UV}^{2\lambda-1}\, 
                      \left( 
                             \frac{q}{2}
                      \right)^{2\lambda}\, 
                      \int_0^\infty z\, U(z)\, K_\lambda(qz)^2\, dz ~.    
\label{estimate}
\end{eqnarray}
For large and small momenta $q$, the above integral is dominated by the 
contributions of the interaction potential $U(z)$ in the regions of 
$z\to 0$ and $z\to \infty$ respectively, where it behaves as in 
\refeq{UVbehav} and \refeq{longrange}. Thus in these regimes we have
\be
{{\cal F}_1 (q)} \sim q^{2\Delta-d}~,
\ee
with the relevant $\Delta$ in the two cases. This behavior is consistent with
field theory expectations as can be seen by performing the integral
transformation \refeq{hankeltransf}. In fact this gives for the leading
contribution to the $c$-function  $c(x) \sim x^{2(d-\Delta)}$.
The results of \cite{Anselmi:2000fu,Anselmi:1996mq} state that generically in
field theory the anomalous dimension $h_*$ at the critical point of an RG flow
is related to the $c$-function as
\be
h_* & = & -\lim_* \frac{\ddot{c}}{\dot{c}}~,
\ee  
where the dot denotes the logarithmic derivative. Using this, we find
\be
h_{\mathrm{UV}/\mathrm{IR}} & = &  \Delta_{\mathrm{UV}/\mathrm{IR}}-d~.
\ee
This result was checked in the UV in \cite{Anselmi:2000fu,Porrati:2001nb}
for some specific examples. In fact we have seen that it should hold for
any holographic flow near the UV and IR fixed points.
To show that the $c$-function is monotonous  along the flow and that the IR fixed
point value is proportional to  $R_\mathrm{IR}^{d-1}$ seems to involve a more
detailed knowledge of the full potential $U(z)$. We will be able to check these
properties in some toy model in the next section.

\section{Models}
\label{astutissimi}

In the previous section we have shown that the TT graviton fluctuation
of an interpolating flow is equivalent to a scattering problem, with
well-behaved QM potentials. In order to make further progress, 
we need to consider some specific model. 
We can deal with a feasible problem if we 
consider QM potentials that are piecewise solvable, and match them together
at an arbitrary point.
In the following we examine two models which correspond
to $\delta$-function and step barrier interactions. 
One can regard them as the most crude approximation to the generic 
potentials. Nevertheless, they turn out to encode most of the generic features
expected from an interpolating flow.\\   
%%%%%%%%%%%%%%%%%%%%%%%%%%%%%%%%%%%%%%%%%%%%%%%%%
\parbox{\textwidth}{\vspace{0.2cm}
%%%%%%%%%%%%%%%%%%%%%%%%%%%%%%%%%%%%%%%%%%%%%%%%%
\parbox{0.5\textwidth}
{
  \refstepcounter{figure}
  \label{thinpen}
  \begin{center}
  \begin{turn}{0}
  \makebox[5cm]{
     \epsfxsize=2.5cm
     \epsfysize=3.5cm
     \epsfbox{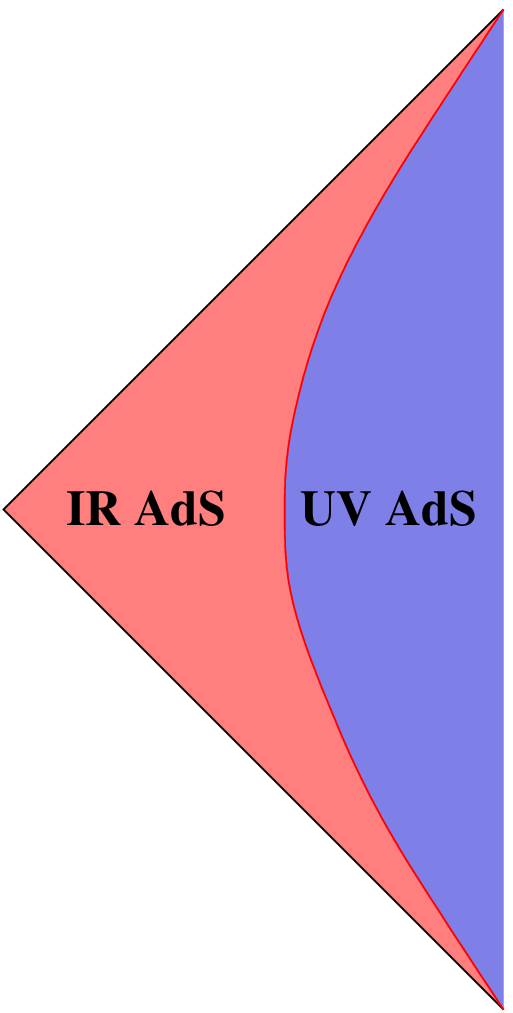}
     }
  \end{turn}
  \end{center}
  \center{{\bf Fig.{\thefigure}.} Thin AdS/AdS wall.}
}\hfill
\parbox{0.5\textwidth}
{
  \refstepcounter{figure}
  \label{thickpen}
  \begin{center}
  \begin{turn}{0}
  \makebox[5cm]{
     \epsfxsize=2.7cm
     \epsfysize=3.5cm
     \epsfbox{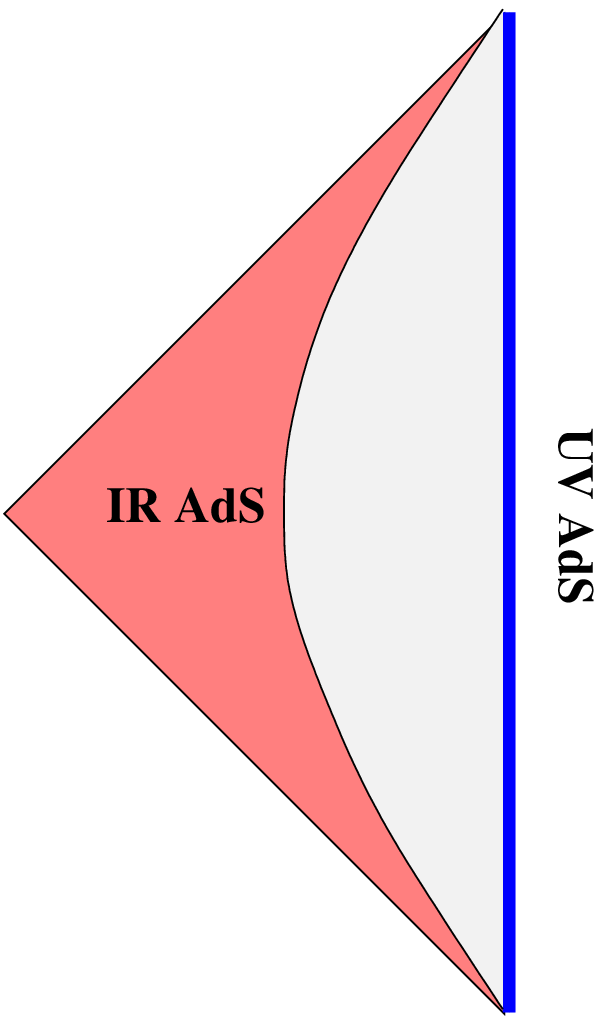}
  }
  \end{turn}
  \end{center}
  \center{{\bf Fig.{\thefigure}.} AdS/IK-space wall.}
}
\center{Schematic diagrams of the models.}
%%%%%%%%%%%%%%%%%%%%%%%%%%%%%%%%%%%%%%%%%%%%%%%%%
}
%%%%%%%%%%%%%%%%%%%%%%%%%%%%%%%%%%%%%%%%%%%%%%%%%

\subsection{Thin AdS/AdS wall}

We construct the simplest interpolating ``flow'' by  gluing
together two AdS spaces with different radii. This is a thin-wall
approximation adapted to the present problem. 
The resulting interaction potential is a $\delta$-function at the gluing
point.

In the $z$-coordinate system (the same conclusion can be obtained 
for instance in the $r$-coordinate) we consider two patches with local 
coordinates $z_1$ and $z_2$ and metrics
\begin{eqnarray}
   \dd s^2_1 ~=~ \frac{R_\mathrm{UV}^2}{z_1^2}\,
                 (\,\dd z_1^2 \,+\, \eta_{ij}\dd x^i\dd x^j\,) 
   \qquad 
   \dd s^2_2 ~=~ \frac{R_\mathrm{IR}^2}{z_2^2}\,
                 (\,\dd z_2^2 \,+\, \eta_{ij}\dd x^i\dd x^j\,)~.
\end{eqnarray}
Requiring that the metric on a constant $z$ hypersurface be continuous
at the gluing point, implies that $\bar{z}_1 R_\mathrm{IR}=\bar{z}_2
R_\mathrm{UV}$. 
In order to impose the correct boundary conditions for the 
fluctuation at the matching
point, it is more convenient to consider the physical 
fluctuation $\chi$ instead of
the rescaled function $\psi$. The equation for $\chi$ reads
\begin{eqnarray}\label{eq:chi}
   \chi'' (z)~+~ (d-1)A'(z)\chi'(z)- k^2 \chi(z)  & = & 0~,
\end{eqnarray}
the solutions of which, in each of the two regions, are the free ones
\begin{eqnarray}\label{Chi_Ex1}
    \chi (z) & = &
    \left(\, \frac{z}{R}\, \right)^\lambda
    \Big[\,
             a\,I_{\lambda}(q\,z) ~+~ 
             b\,K_{\lambda}(q\,z)\,
    \Big]~,
\end{eqnarray}
with $a\,=\,0$ in the IR. The appropriate matching conditions 
on $\chi$ are
\begin{eqnarray}
\label{matching}
     \chi_\mathrm{UV}  (\bar{z}_1)  ~=~  \chi_\mathrm{IR} (\bar{z}_2)\hspace{1.75cm}
     \chi_\mathrm{UV}' (\bar{z}_1)  ~=~  \chi_\mathrm{IR}'(\bar{z}_2)~.
\end{eqnarray}
The first condition follows from requiring continuity of the metric, while 
the second one is a consequence of integrating \refeq{eq:chi} across the
gluing surface. 

The flux is now easily obtained, computing the two contributions to
\refeq{firstresult}. 
In this case the potential vanishes at the origin, so that the
logarithmic contribution is the pure UV fixed point term
\begin{eqnarray}
\label{pureUV}
{\cal F}_\mathrm{log}(q) & = & \underbrace{
                               -\,\frac{4\lambda\, (-1)^{\lambda}}
                                       {\Gamma(\lambda)\,\Gamma(\lambda+1)}\,
                                  R_\mathrm{UV}^{2\lambda-1}\,
                                  \left(\frac{q}{2}\right)^{2\lambda}\log\,q
                             }_{
                                 {\cal F}_\mathrm{CFT}^\mathrm{UV}(q)
                             }~+~{\cal O}(q^{2\lambda})~.
\end{eqnarray}
The non trivial part of the flux ${\cal F}_\mathrm{int}$ in 
eq.~(\ref{firstresult}) reduces here to the quotient of $a_\mathrm{UV}$ 
and $b_\mathrm{UV}$ as follows
\begin{eqnarray}\label{Interaction}
    {\cal F}_\mathrm{int}(q)
    &=& -\,\frac{4\lambda}{\Gamma(\lambda)\,\Gamma(\lambda+1)}\,
        R_\mathrm{UV}^{2\lambda-1}\left(\frac{q}{2}\right)^{2\lambda}\,
        \frac{
                a_\mathrm{UV}
             }
             {
                b_\mathrm{UV}
             }
\end{eqnarray}
with
\begin{eqnarray*}
    \frac{
           a_\mathrm{UV}
         }
         {
           b_\mathrm{UV}
         } 
    = \frac{
             \Big[\,
                      K_{\lambda+1}(q\bar{z}_1r)K_{\lambda}(q\bar{z}_1)-
                      K_{\lambda}(q\bar{z}_1r)K_{\lambda+1}(q\bar{z}_1)\,
             \Big]q\bar{z}_1 r 
             + d\left(r-1\right)
             K_{\lambda}(q\bar{z}_1r)K_{\lambda}(q\bar{z}_1)
           }
           {
             \Big[\,
                      K_{\lambda+1}(q\bar{z}_1r)I_{\lambda}(q\bar{z}_1)+
                      K_{\lambda}(q\bar{z}_1r)I_{\lambda+1}(q\bar{z}_1)\,
             \Big]\,q\bar{z}_1 r 
             + d\left(r-1\right)
             K_{\lambda}(q\bar{z}_1r)I_{\lambda}(q\bar{z}_1)
           } 
\end{eqnarray*}
where we have defined the quotient of the AdS radii 
$r=R_\mathrm{IR}/R_\mathrm{UV}$. From the above expression it is evident
that ${\cal F}_\mathrm{int}$  vanishes if $R_\mathrm{IR}=R_\mathrm{UV}$,
and one recovers the pure AdS result from \refeq{pureUV}. One can also, 
check using asymptotic expansions of the Bessel functions, that ${\cal
F}_\mathrm{int}$ vanishes in the large $q$ limit. 

To check the IR behavior of the flux, we expand $a_\mathrm{UV}/b_\mathrm{UV}$
for small $q$
\begin{eqnarray}
\label{fluxIR}
     \frac{a_\mathrm{UV}}{b_\mathrm{UV}} 
     =  (-1)^{\lambda}\,
          \frac{
                 R_\mathrm{IR}^{d-1}-R_\mathrm{UV}^{d-1}
               }
               {
                 R_\mathrm{UV}^{d-1}
               }\,\log q 
     + \frac{\Gamma(\lambda)\Gamma(\lambda-1)}{2}\,
         \frac{R_\mathrm{IR}-R_\mathrm{UV}}{R_\mathrm{UV}}\,
         \left(\,
                  \frac{2}{q\bar{z}_1}
         \right)^{d-2}
     +\ldots
\end{eqnarray}
where dots stand for higher order and contact terms. Notice that we have
included the lowest order analytic term of the flux, which is quadratic
in momentum and deserves a special attention.
Inserting \refeq{fluxIR} into 
\refeq{firstresult}, we get the following small $q$ expansion
of the flux
\begin{eqnarray}
\label{smallq}
    {\cal F}(q) &=& {\cal F}_\mathrm{CFT}^\mathrm{IR}(q)
                ~+~ \frac{1}{2\,(\lambda-1)}\,
                    \left(\frac{R_{\mathrm{UV}}}{\bar{z}_1}\right)^{d-2}\,
                    (R_\mathrm{UV}-R_\mathrm{IR})\,q^2\,
                    ~+~ \ldots
\end{eqnarray}
Notice that the leading non-analytic term of eq.~(\ref{Interaction})
has combined exactly with ${\cal F}_\mathrm{CFT}^\mathrm{UV}$ to reproduce 
the pure AdS IR result, independently of $\bar{z}_1$. 

The last decisive test that we have to perform on the supposed $c$-function, 
is to check that it is in fact positive and monotonic. 
An analytical evaluation of the integral in \refeq{hankeltransf} is not
possible, therefore we have computed it numerically, for some specific values
of the parameters.
Note that the quadratic term in the small $q$ expansion
\refeq{smallq} is not a contact term and at large distances gives 
a contribution $\propto x^{d-2}$ in \refeq{hankeltransf}. An essential step for obtaining a
meaningful central function is to subtract this contribution. 
Namely, we have computed the following ``renormalized'' $c$-function
\begin{eqnarray}
   \hat{c}(x) &=& c(x) ~-~ (4\pi)^{d/2}\,
                         \frac{\Gamma(\lambda-1)}{8\,(\lambda-1)}\,
                         \left(\,
                                  \frac{R_\mathrm{UV}}{\bar{z}_1}\,
                         \right)^{d-2}\,
                         (\,R_\mathrm{UV}-R_\mathrm{IR}\,)\,
                         x^{d-2}~.
\end{eqnarray}
A plot of $\hat{c}(x)$ is drawn in Fig. \ref{cren1} below, where we have 
normalized $\hat{c}(0)$ to $1$.\\
%%%%%%%%%%%%%%%%%%%%%%%%%%%%%%%%%%%%%%%%%%%%%%%%%
\parbox{\textwidth}{\vspace{0.2cm}
%%%%%%%%%%%%%%%%%%%%%%%%%%%%%%%%%%%%%%%%%%%%%%%%%
  \refstepcounter{figure}
  \label{cren1}
  \begin{center}
  \begin{turn}{0}
  \makebox[5cm]{
     \epsfxsize=5cm
     \epsfysize=5cm
     \epsfbox{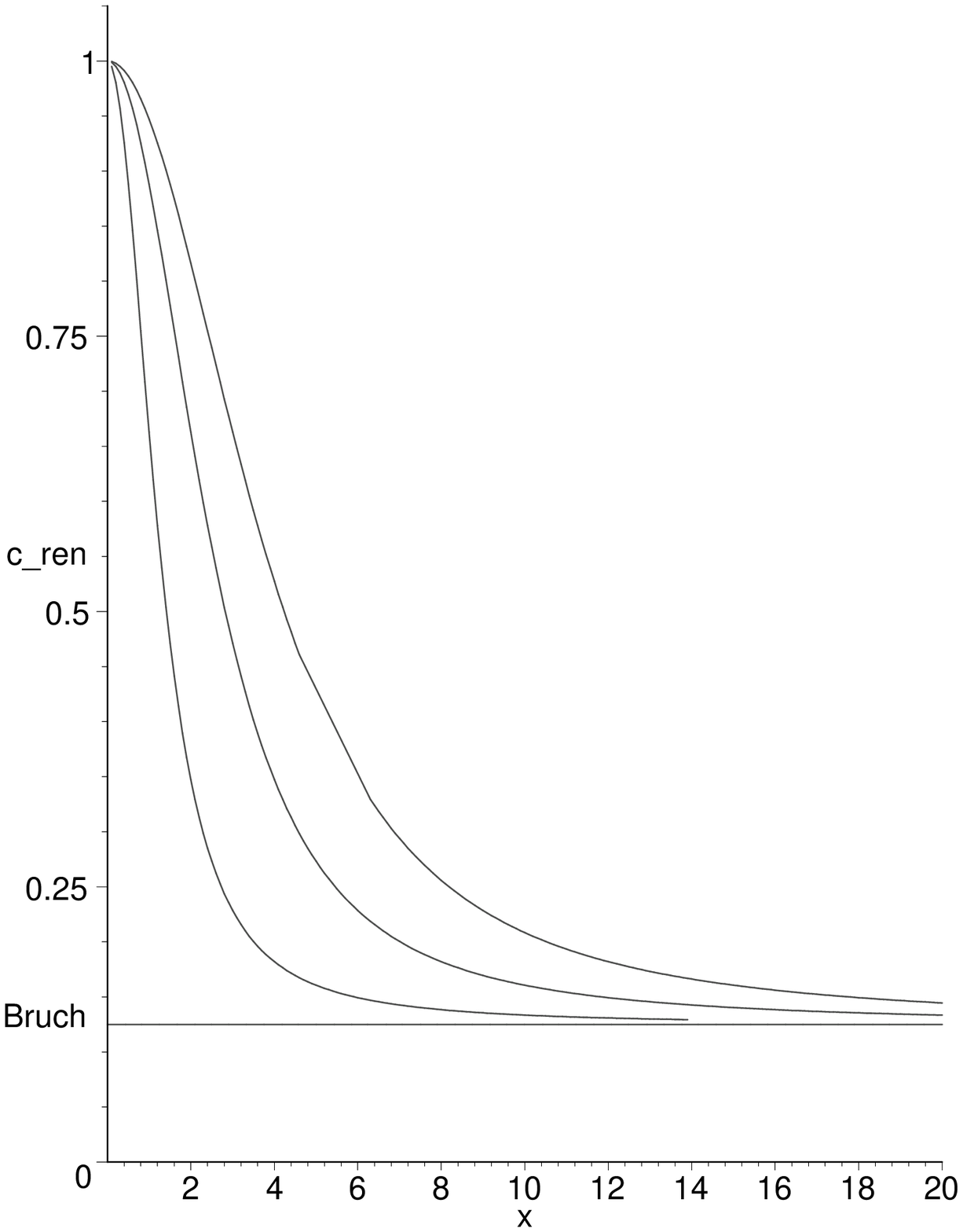}
  }
  \end{turn}
  \end{center}
  \center{{\bf Fig.{\thefigure}.} The $c$-function 
                                  $\hat{c}(x)$ for $d=4$,\\
                                  $R_\mathrm{UV}=2$, $R_\mathrm{IR}=1$
                                  and $\bar{z}_1=1$, $2$, $3$.}
%%%%%%%%%%%%%%%%%%%%%%%%%%%%%%%%%%%%%%%%%%%%%%%%%
}\\
%%%%%%%%%%%%%%%%%%%%%%%%%%%%%%%%%%%%%%%%%%%%%%%%%

\noindent
We see that 
$\hat{c}(x)$ obeys the properties expected of a central function, and
in particular decreases towards $c_\mathrm{IR}>0$ at large distances. This
property has not been checked before in the context of holographic flows.

Let us comment on the issue of the quadratic term. 
In the context of holographic flows these terms have been discussed previously
in \cite{Bianchi:2001de}. In this reference it is pointed out that 
$q^2$-terms in the effective boundary action lead to   
massless poles in the correlation functions. Accordingly, it turns out  
that in the flow of \cite{Girardello:2000bd} such terms cancel, which is 
consistent with 
the interpretation of an IR confining theory, with a mass-gap. On the 
other hand, a massless pole in a Coulomb branch flow is expected, as arising
from the Goldstone boson associated with broken conformal invariance induced by
the non-zero vev.
It is more difficult to give a clear interpretation to our present model, since
it does not arise as a smooth flow, and in particular there is no obvious 
description in terms of small perturbation of the UV CFT. Nonetheless, the 
resulting $c$-function is certainly suggestive of a non-trivial boundary 
theory, which has a UV {\em and} an IR CFT fixed points. 

At the end of section \ref{scatt} we have analyzed the small $q$ behaviour 
of the flux stemming from an interaction potential 
characteristic of a smooth interpolating flow. We have seen that generically
the first correction to the (logarithmic) fixed point behaviour goes like
$q^{2\Delta_\mathrm{IR}-d}$. The analysis of section \ref{scatt} relied on the
expansions \refeq{asymphi} \refeq{asymA} which are not valid 
for a delta-function potential, and as a result it gives rise to a spurious
IR contribution which is quadratic. A generic long-range potential which
arises from a smooth flow will prevent such quadratic terms.

\subsection{AdS/IK-space domain wall}

The previous model encodes most of the expected features related to the 
$c$-function of an interpolating flow. Nevertheless one might think that 
the result may lack generic features. Here we want to discuss an improved 
construction. Roughly speaking, we will give a finite thickness to the wall 
separating two different AdS regions, thus simulating a setting closer in 
spirit to an interpolating flow.
In the language of the Schr\"odinger problem we are going to consider a 
step potential 
\begin{eqnarray}
     U &=& \left\{
                   \begin{array}{ll}
                     \omega^2  & {\rm \,in\;\;patch\;}\; \{z_1\} \\
                        0      & {\rm \,in\;\;patch\;}\; \{z_2\} \
                   \end{array}
           \right.
\end{eqnarray} 
combined with an additional $\delta$-function located at the gluing point, 
{\em i.e.} we match to an AdS space (in the IR) with warp factor 
$A_{II}(z_2)\,=\,\log({R}_{\rm IR}/z_2)$ a spacetime described by 
the a warp factor $A_{I}(z_1)$ obeying the equation
\begin{eqnarray}
   \left(\frac{d-1}{2}\right)^2\,\left(\frac{dA_{I}}{dz_1}\right)^2 ~+~ 
   \frac{d-1}{2}\,\frac{d^2A_{I}}{dz_1^2} & = &
   \frac{\lambda^2-1/4}{z_1^2}\,+\,\omega^2~.
\end{eqnarray}
The general solution to this equation is 
\begin{eqnarray}\label{strangewarp}
 A_{I}(z_1) & = & \frac{2}{d-1}\log 
              \Big[\,
                       k_1\sqrt{z_1}\,K_\lambda (\omega z_1) \,+\,
                       k_2\sqrt{z_1}\,I_\lambda (\omega z_1)\,
              \Big]~,
\end{eqnarray}
where $k_1$ and $k_2$ are two positive constants,
and we call the resulting spacetime an ``IK-space'', for obvious reasons.
Given the warp factor for the metric, it is in principle straightforward 
to obtain the solution for $\phi(z)$ from \refeq{BPS},
and eventually the supergravity 
potential $V$. However, for the computation of the $c$-function this information
is not needed, and we have not performed this analysis here.
In the limit $z_1\to 0$ the warp factor asymptotes to 
\begin{eqnarray}
   A_{I}(z_1) & \to & \log
       \left[\,
                \frac{
                       {R}_{\rm UV}
                     }
                     {
                       z_1
                     }
       \right]~\quad\mathrm{with}\quad
    \quad\quad
    {R}_{\rm UV}^{2\lambda-1} \,=\, 
                            \left(\,\frac{2}{\omega}\,\right)^{2\lambda}
                            \left(\,
                                    \frac{\Gamma(\lambda)}{2}\,
                            \right)^{2} 
                             k_1^2
\end{eqnarray}
and so the space is asymptotically AdS.  The solution of the fluctuation 
equation (\ref{eq:chi}) in the patch $\{z_1\}$ is given by
\begin{eqnarray}\label{PerturbedBessel}
 \chi_{I}(z_1) &=& \frac{
                           a_{\rm UV}\,
                           I_{\lambda}(\epsilon\,z_1)
                           ~+~
                           b_{\rm UV}\,
                           K_{\lambda}(\epsilon\,z_1)
                        }
                        {
                           k_1\,K_{\lambda}(\omega\,z_1)
                           ~+~
                           k_2\,I_{\lambda}(\omega\,z_1)
                        }~,
  \quad\quad\quad
  \epsilon\,=\,\sqrt{\omega^2+q^2}
\end{eqnarray}
and $\chi_{II}(z_2)$ is the free solution (\ref{Chi_Ex1}) used in the 
first model. 

Notice that in principle one can choose the gluing point $\bar{z}_1$ such that $A_{I}(z_1)$ is monotonic, 
and tune the integration constant $k_2$ 
so to match the two regions smoothly, {\em i.e.} with continuous derivatives
of the warp factors. We make here a
simplifying assumption and set $k_2\,=\,0$. The requirement 
of continuity of the metric relates the two patches to each other via
\begin{eqnarray}
   \bar{z}_2 &=& \frac{{R}_{\rm IR}}{{R}_{\rm UV}}\,
             \left(\,
                     \frac{2^{\lambda-1}\,\Gamma(\lambda)}{\omega^{\lambda}\,\sqrt{\bar{z}_1}\,K_{\lambda}(\omega\,\bar{z}_1)}\,
             \right)^{\frac{1}{\lambda-1/2}}~.
\end{eqnarray}   
For $\omega\to 0$ this approaches exactly the corresponding  relation 
in the first example.

Requiring $\chi_{I/II}(z_{1/2})$ to be smooth as in  \refeq{matching},
allows to determine the quotient of $a_{\rm UV}$ and $b_{\rm UV}$ as in
the previous section. To shorten somewhat the expression, we have not 
written out explicitly the derivatives of Bessel functions, denoted with
a prime
{\small
\begin{eqnarray*}
   \frac{a_{\rm UV}}{b_{\rm UV}}
   = \frac{
              \left[\,
                          \frac{d}{\bar{z}_2}\,
                          K_{\lambda}(q\,\bar{z}_2)
                          -q\,K_{\lambda+1}(q\,\bar{z}_2)\,
              \right]\,
              K_{\lambda}(\epsilon\,\bar{z}_1)
              -{
                     [K_{\lambda}(\epsilon\,\bar{z}_1)]'
                     -K_{\lambda}(\epsilon\,\bar{z}_1)\,
                     \frac{[K_{\lambda}(\omega\,\bar{z}_1)]'}
                     {K_{\lambda}(\omega\,\bar{z}_1)}
                   }
                        \,K_{\lambda}(q\,\bar{z}_2)
            }
            {
              \left[\,
                          \frac{d}{\bar{z}_2}\,
                          K_{\lambda}(q\,\bar{z}_2)
                          -q\,K_{\lambda+1}(q\,\bar{z}_2)\,
              \right]\,
              I_{\lambda}(\epsilon\,\bar{z}_1)
              -{
                     [I_{\lambda}(\epsilon\,\bar{z}_1)]'
                     -I_{\lambda}(\epsilon\,\bar{z}_1)\,
                     \frac{[K_{\lambda}(\omega\,\bar{z}_1)]'}
                     {K_{\lambda}(\omega\,\bar{z}_1)}                           
                                      }\,
                   K_{\lambda}(q\,\bar{z}_2)
            }
\end{eqnarray*}
}\\[-1ex]
The quotient of Jost functions can be worked out and is proportional 
to the quotient of $a_{\rm UV}$ and $b_{\rm UV}$ as  given below
\begin{eqnarray}
\label{fluxintIK}
   {\cal F}_{\rm int}(q)
   &=& -\,\frac{4\lambda}{\Gamma(\lambda)\,\Gamma(\lambda+1)}\,
       R_{\rm UV}^{2\lambda-1}\,
       \left(\,\frac{\omega^2+q^2}{4}\,\right)^{\lambda}\,
       \frac{a_{\rm UV}}{b_{\rm UV}}~.
\end{eqnarray}
In this case, however, the interaction potential does not vanish at the 
origin, hence the solutions (\ref{PerturbedBessel}) do not behave as the free 
Bessel solutions as mentioned in section \ref{scatt}. Now the contribution 
from the singular solution is not the pure AdS UV term and we obtain instead
\begin{eqnarray}
\label{irrecontr}
{\cal F}_\mathrm{log}(q) & = &
                               -\,\frac{4\lambda\, (-1)^{\lambda}}
                                       {\Gamma(\lambda)\,\Gamma(\lambda+1)}\,
                                  R_\mathrm{UV}^{2\lambda-1}\,
                                  \left(\frac{q^2+\omega^2}{4}\right)^
                                  {\lambda}\log\sqrt{q^2+\omega^2} 
                         ~+~ {\mathcal{O}}(q^0)
\end{eqnarray}
This reduces to ${\cal F}_{\rm CFT}^{\rm UV}$  in the large $q$ limit. 
However, since \refeq{irrecontr} is analytic at small $q$, the correct 
non-analytic term in the IR limit must follow from  
${\cal F}_\mathrm{int}(q)$ alone. Making the expansion and 
utilizing some Bessel function identities, one can extract the leading
non analytic contribution of  \refeq{fluxintIK} together with its leading 
quadratic term, written for $d\,=\,4$ below
\begin{eqnarray}\label{Letzte}
   {\cal F}_{\rm int}(q) &=& \underbrace{
                                         - \frac{1}{4}\,R_{\rm IR}^3
                                           \,q^4\,\log\,q 
                                        }_{
                                           {\cal F}_\mathrm{CFT}^\mathrm{IR}(q) 
                                        }\\
                         &+&  R_{\rm UV}^{3}\,
                        \left[\vbox{\vspace{3ex}}\right.\,
                        \frac{1}{8}\,\omega^4\,\bar{z}_1^2\,
                        K_1(\omega\,\bar{z}_1)\,
                        K_3(\omega\,\bar{z}_1)
                  ~-~ \frac{1}{2}\,
                            \frac{
                                   \bar{z}_1\,+\,\bar{z}_2
                                 }{\bar{z}_2^3}\,
                        \left(
                               \frac{{R}_{\rm IR}}{{R}_{\rm UV}}
                        \right)^3  
                        \left.\vbox{\vspace{3ex}}\right]\,q^2                  
                  ~+~\ldots\nonumber  
\end{eqnarray}
That we find the correct IR term is a non trivial check of the model 
and in agreement with the expectations. 
One can work out the quadratic terms, too. Now there are two sources of them, 
eq.~\refeq{irrecontr} and eq.~\refeq{Letzte}. In the limit $\omega\to 0$  
the quadratic terms reproduce precisely the terms of the first example. 
We expect that  after proper subtractions we should find a function $\hat{c}$ 
which interpolates the two central charges correctly. 
In this case the numerical evaluation of the integral in 
eq.~\refeq{hankeltransf} seems to be problematic and the corresponding 
plot is still missing.

\section{Conclusions}

In this paper, we have addressed the issue of holographic interpolating flows.
Contrary to Coulomb branch flows and IR-confining flows, in the literature
there were no explicit examples of interpolating, {\em i.e.} IR-conformal,
flows. Therefore the test of the AdS/CFT correspondence in these settings
has been somewhat limited.
In the first part we have characterized such flows in terms of their
superpotentials and their expansions near the UV and IR fixed points.
We have shown that the language of scattering theory is suitable for 
studying the fluctuations of the transverse traceless part of the metric, 
which are relevant for the computation of a holographic $c$-function.

Following these lines, we have proposed two explicit models which realize
interpolating flows. The first model consists of a background space-time
in which two AdS regions are glued together at an arbitrary distance from
the boundary. In the second model we have replaced the outer AdS space 
(UV region) with an asymptotically AdS space. The corresponding scattering 
problems are solved explicitly.

In both cases we have shown that the $c$-function is non-trivial, and
interpolates correctly between the UV and IR fixed point central charges.
In particular, in our models the holographic $c$-function defined in terms 
of the superpotential \cite{Freedman:1999gp} is trivial, and the 
related beta-function is ill-defined. On the other hand there is no 
problem to define a ``proper'' beta-function which is defined in 
term of the $c$-function \cite{Anselmi:2000fu,Anselmi:1999xk}.

In the models we considered, we have found a quadratic term in the 
small momentum expansion of the flux-factor. We gave an argument that 
such a term will not be present in a smooth deformation flow. Therefore
this feature should be an artefact of the toy models considered. Especially
in the thin AdS/AdS domain wall, this should be related to the holographic 
image of a ``brane'' sitting in the bulk. It would be interesting to make 
this picture more clear.

\vskip .8cm

\noindent
{\bf Note added:} A paper on a related subject, investigating
                  a flow in two dimensional CFT, appeared on the arXive on 
                  the same day \cite{Berg:2001ty}.

\section*{Acknowledgements} 

We wish to thank G. Ferretti, H.~S. Reall, and D. Waldram 
for discussions and helpful comments, and A.~C. Petkou for e-mail
correspondence. We are grateful to W. M\"uck for e-mail correspondence
and comments on the manuscript.
This work has been partially supported by PPARC through SPG\#613.

\appendix

\section{Potential scattering}
\label{complexj}

We follow reference \cite{regge}.
The radial Schr\"odinger equation for the $l$-th partial wave for 
scattering off a central potential is  
\be
-\frac{d^2}{dz^2}\psi (z)+\left[ \frac{l(l+1)}{z^2}+U(z)-k^2\right]\psi (z) =
0~. 
\label{schro}
\ee 
Let us define $\lambda=l+1/2$. One can conveniently define two sets of
independent solutions for the equation \refeq{schro}, which refer to boundary
conditions given respectively at $z\to 0$ and $z\to \infty$. 
The ``{\em regular}'' solutions are defined by the 
boundary condition
\be
\lim_{z\to 0} \varphi (\pm\lambda,k,z)z^{\mp\lambda-1/2} & = & 1~,
\label{defreg}
\ee
so that the non-singular and singular solutions are conveniently labeled by the
sign in front of $\lambda$. 
The two solutions are always linearly independent
(except for $\lambda=0$). 
The {\em Jost solutions} are defined by the boundary condition
\be
\lim_{z\to \infty}  \mathrm{f}(\lambda,\pm k,z)\e^{\pm ikz} & = & 1~.
\label{defjost}
\ee
For the application in mind we are interested in
$\mathrm{f}(\lambda,-k,z)$, since it falls off exponentially at infinity upon
Wick rotation to the positive imaginary axis. 
The two sets of solutions are related by 
\be
\mathrm{f} (\lambda,\pm k,z) & = & \frac{1}{2\lambda}\left[ 
\jost (\lambda,\pm k) \varphi (-\lambda,\pm k,z)- \jost (-\lambda,\pm k)
\varphi (\lambda,\pm k,z)\right]\label{newbasis}\\
\varphi (\pm \lambda,k,z) & = & \frac{1}{2ik}\left[ 
\jost (\pm \lambda,k)  \mathrm{f}(\lambda,-k,z)- \jost (\pm\lambda,-k)
\mathrm{f} (\lambda,k,z)\right]~.
\ee
The coefficients $\jost(\pm\lambda,\pm k)$ are called {\em Jost functions} 
and can be obtained by suitable Wronskians, {\em e.g.}
\be
  \jost (\lambda,k) 
  & \equiv & W[\mathrm{f}(\lambda,k,z),\varphi (\lambda,k,z)]~.
\ee
They are independent of $z$, since the Wronskian of any two 
solutions of \refeq{schro} is a constant. We now specialize to 
Euclidean momenta $q=i|k|$.
The regular solutions and the Jost solutions for the free case, {\it i.e.} 
$\,U(z)\,\equiv\,0\,$, read respectively
\begin{align}
   \varphi_0 (+\lambda_,q,z)&~=~ 
   2^\lambda\,\Gamma(\lambda+1)\,q^{-\lambda}\,\sqrt{z}\,I_\lambda (qz)~, &
   f_0(\lambda,-q,z)~=~&\sqrt{2qz/\pi}\,K_\lambda (qz) \\
   \varphi_0(-\lambda,q,z) &~=~ \frac{2^{1-\lambda}\,q^\lambda}
                                {\Gamma(\lambda)}\,\sqrt{z}\,K_\lambda (qz)~, &
   f_0(\lambda,+q,z)~=~&\sqrt{2\,\pi\,qz}\,I_\lambda (qz)
\end{align}  
and the free Jost functions are
\be
    \jost_0 (\lambda,-q) ~=~  \sqrt{2/\pi}\,2^\lambda\,
                              \Gamma(\lambda+1)\,q^{-\lambda+1/2}~, 
    \qquad & \qquad \jost_0 (-\lambda,-q) ~=~ 0~.
\ee 
The Jost solution can be constructed recursively, {\it i.e.} 
$f(\lambda,\pm q,z)= \sum_{n=0}^\infty f_n(\lambda,\pm q,z)$. 
Each term in this expansion can be obtained by
\be
   f_n(\lambda,\pm q,z) & = & \int_z^\infty\,B(\lambda,q,z,\xi)\,U(\xi)\, 
   f_{n-1}(\lambda,\pm q,\xi)\,d\xi~.
\ee
Here the kernel is $B(\lambda,q,z,\xi)\,=\,\sqrt{z\xi}\,\left[\,
K_\lambda (q\xi)\,I_\lambda (qz)\,-\,K_\lambda (qz)\,I_\lambda (q\xi)\,
\right]$. The integral representation for the Jost function reads
\be
\jost (\pm \lambda, -q) &=& \jost_0 (\pm \lambda,-q) 
                        ~+~ \int_0^\infty\,
                                     \varphi_0 (\pm\lambda,q,\xi)\,
                                     U(\xi)\,f(\lambda,-q,\xi)\,d\xi~.
\label{intejostfun}
\ee
These expressions are useful for deriving Born type expansions. 
In particular from eq.~\refeq{intejostfun} we obtain an expansion of 
the Jost function 
\be
\jost (-\lambda,-q) & = & \sum^\infty_{n=0}\jost_n (-\lambda,-q)~,
\label{bornjost}
\ee
whose first order correction reads
\be
%   \jost_1 (+\lambda,-q)  
%    &=& \sqrt{2/\pi}\,2^\lambda\,\Gamma (\lambda+1)\,
%         q^{-\lambda+1/2}\,\int^\infty_0 \xi\,U(\xi)\,I_\lambda\,(q\xi)\,
%         K_\lambda (q\xi)\,d\xi\nonumber\\
   \jost_1 (-\lambda,-q) 
    &=& \sqrt{2/\pi}\,\frac{2^{-\lambda}}{ \Gamma (\lambda)}\,
        q^{\lambda+1/2}\,\int^\infty_0 \xi\, U(\xi) \, 
        K_\lambda^2 (q\xi)\,d\xi~.
\label{jost1}
\ee
We refer the reader to \cite{regge} for further details.

\section{Counterterms and anomalies}
\label{reno}

In this Appendix we make some comments on the expansions of the superpotentials
of section \ref{superpexp} in relation to the issue of ``holographic 
renormalization'' \cite{deHaro:2001xn,Kalkkinen:2001vg,Bianchi:2001de}.
 
In general the on-shell action, which is used for computing holographic
correlation functions \cite{Gubser:1998bc,witten}, suffers from divergences
arising integrating certain terms near the AdS boundary. These
divergences can be isolated and subtracted, in order to obtain a finite
renormalized on-shell action. Some of these terms may be determined by analyzing
the simple
background solutions, though more general solutions ($x$-dependent) must be 
considered to find all of the counterterms. Here we focus our attention
on the former type, so that all possible divergent terms are contained in the
boundary term
\be
S & = & \int_{z=\epsilon} d^d x \sqrt{\hat{g}} W(\phi)~, 
\label{Sonshell}
\ee 
where $\epsilon $ is a (UV or IR) cut-off, and $\hat{g}_{ij}$ is the
induced metric. Given the expansions \refeq{asymphi} and \refeq{asymA}, for a
generic background flow near a fixed point, divergences can arise from the
leading terms in the expansion\footnote{For compactness of notation in this
Appendix we have set $R=1$.}
\be
\sqrt{\hat{g}} W(\Phi) & = & \frac {1} {\epsilon^d} \left(W_0+\frac{1}{2}
W_0^{(2)}\epsilon^{2(d-\Delta)}+\cdots\right)~.
\label{Sdiv}
\ee  
First, let us rule out the possibility that divergences (hence anomalies) 
arise near the IR fixed point. According to \refeq{boundIR}, the
exponents $n(d-\Delta)-d$ are negative, and therefore there are no 
divergences as the IR region is reached at $\epsilon\to \infty$. 

Near the UV boundary, the divergent terms are those of order 
$n<\left[d/(d-\Delta)\right]\,$.
A finite term can arise only for operators saturating the above bound. We have
seen that they give rise to the logarithmic-modified expansion for $W$
\refeq{Wlog}. In turn, the expansions \refeq{asymphi} and \refeq{asymA} become
more complicated. However, at leading order they are unchanged, so that the
divergent part of \refeq{Sonshell} can be read off as
\be
\Big[\sqrt{\hat{g}} W(\Phi)\Big]_\mathrm{div} & = & \frac{1}{\epsilon^d}W_0+
\cdots+\frac{W_0^{(n)}}{n!}
+(d-\Delta)\frac{X_0^{(n)}}{n!}\log\epsilon~.
\ee  
The first $n-1$ terms are divergent and must be subtracted from the on-shell 
action. We have seen that the finite term was left undetermined by the equations
of motion, and here we see explicitly that it corresponds to a choice of
renormalization scheme. 
Furthermore we can now interpret the logarithmic term. Upon
varying the on-shell action this contributes to the fixed point conformal
anomaly via the mechanism of \cite{Henningson:1998gx}. Thus,
as expected \cite{deHaro:2001xn,Petkou:1999fv},
we get the following contribution to the Weyl anomaly due to scalar operators
of dimensions as in \refeq{fractop} 
\be
{\cal A}_\mathrm{matter} & = & \frac{d}{(n+1)!} X_0^{(n)}\Phi^n~. 
\ee
Interestingly, it encodes information about the gravity potential $V$, up to
order $n$.

\section{Coefficients of the asymptotic expansions}
\label{coefficients}

Here we list few coefficients in the expansions of $\phi$ and $\e^A$
as indicated in eq.~(\ref{asymphi})
\begin{eqnarray*}
   c_1 &=& \frac{R}{2}\,\frac{W_0^{(3)}}{\Delta-d}\\
   c_2 &=& \frac{R^2}{4}\,\frac{W_0^{(3)}{}^2}{(\Delta-d)^2} ~-~
           \frac{R}{12}\,\frac{W_0^{(4)}}{(\Delta-d)}~+~
           \frac{1}{8}\,
           \frac{(\Delta-d)}{[\,2\,(\Delta-d)-1\,]\,(d-1)}
\end{eqnarray*}
and eq.~(\ref{asymA})
\begin{eqnarray*}
    a_2 &=& ~-~ \frac{1}{4}\,\frac{
                                     (\Delta-d)
                                  }
                                  {
                                     (d-1)\,[\,2\,(\Delta-d)\,-\,1\,]
                                  }\hspace{5.8cm}\\
    a_3 &=& ~-~ \frac{1}{3}\,\frac{
                                      W_0^{(3)}\,R
                                  }
                                  {
                                     (d-1)\,[\,3\,(\Delta-d)\,-\,1\,]
                                  }
\end{eqnarray*}
with
\begin{eqnarray*}
   W_0^{(3)} &=& \frac{R}{3\,\Delta\,-\,2\,d}\,V_0^{(3)}\\
   W_0^{(4)} &=& \frac{R}{4\,\Delta-3\,d}\;
           \left[
                  V_0^{(4)}+\frac{3\,d\,(\Delta-d)^2}{2\,R^2\,(d-1)}-
                  \frac{3\,R^2\,V_0^{(3)}{}^2}{(3\Delta-2d)^2}
                \right]~.
\end{eqnarray*}
The coefficients in the expansion of $V_\mathrm{QM}$ in
eq.~(\ref{sugratoqm})
follow immediately and the first nonzero one reads
\begin{eqnarray*}
     v_{2} &=&  ~-~ \frac{1}{4}\;
                    \frac{
                           (\Delta-d)^2
                         }
                         {
                           2\,(\Delta-d)\,-\,1
                         }\;[\,2\,\Delta-d\,]~.\hspace{1cm}
\end{eqnarray*}
In principle these expansions are valid in the UV and the IR.
However, in the IR, due to the bound of eq.~(\ref{boundIR}) some of the 
denominators can vanish, making the expansion not valid.
This happens for operators of IR
dimension $\Delta_\mathrm{IR}=d+1/n$, for some integer $n$.
In the UV the sign of $v_{2}$ is positive due to the bound of 
eq.~(\ref{boundUV}). 
The first coefficient of the expansion of $\vdqm$ 
in eq.~(\ref{sugratoqm}) is
\begin{eqnarray*}
 \tilde{v}_{2} &=& \frac{1}{4}\;
                   \frac{
                          (\Delta-d)^2
                        }
                        {
                           2\,(\Delta-d)-1
                        }\;[\,2\,\Delta-d\,+\,2(1-d)\,]~.
\end{eqnarray*}

\end{document}